\documentclass[12pt,a4paper]{article}

\usepackage{array}
\usepackage{float}
\usepackage{amsmath}
\usepackage{amsfonts}
\usepackage{amssymb}
\usepackage{latexsym}
\usepackage{epsfig}
\usepackage{graphicx}
\usepackage{mathrsfs}
\usepackage[authoryear]{natbib}
\usepackage[labelsep=colon,font=it,labelfont=sc]{caption}
\usepackage{theorem}
\usepackage{url}
\usepackage{rotating}
\usepackage{multirow}
\usepackage{fancyhdr}
\usepackage{sectsty}
\usepackage{lscape}
\usepackage[truetex]{color}
\usepackage{booktabs}
\usepackage{setspace}
\usepackage[colorlinks=true]{hyperref}
\hypersetup{
	colorlinks = true,
	linkcolor = blue,
	anchorcolor = blue,
	citecolor = blue,
	filecolor = blue,
	urlcolor = blue
}
\usepackage[font = footnotesize,textfont = it]{subcaption}
{
	\theorembodyfont{\upshape} 
	\theoremheaderfont{\scshape} 
	
}
\newtheorem{assumption}{Assumption}

{\theorembodyfont{\upshape} 
	
}

\newtheorem{proposition}{{\sc Proposition}}

\newenvironment{proof}[1][Proof]{\bigskip \noindent \textbf{#1:} }{\  \rule{0.5em}{0.5em}}

\theoremheaderfont{\bfseries}
\theorembodyfont{\upshape}
\sectionfont{\centering \sc}
\subsectionfont{\centering \sc}
\subsubsectionfont{\centering \sc}
\allsectionsfont{\centering \sc}
\allowdisplaybreaks

\bibpunct[:~]{(}{)}{;}{a}{,}{;}
\usepackage[top = 3cm, left = 2.5cm, right = 2.5cm, bottom = 3cm]{geometry}

\begin{document}

\begin{center}
		\textsc{\large GLOBAL FACTORS FOR LOCAL SHOCKS IN A DATA-SCARCE ENVIRONMENT: WITH AN APPLICATION TO REGIONAL FISCAL MULTIPLIERS IN ITALY$^*$}
		\renewcommand{\thefootnote}{}
        
		\footnote{
			\hspace{-7.2mm}
            $^*$We are grateful to Marco Bernardini, Efrem Castelnuovo, Sergio Destefanis, Matteo Fragetta, Raffaella Giacomini, Daniel Lewis, Francesco Simone Lucidi, Camilla Mastromarco, Giuseppe Ragusa, and Massimiliano Tancioni for helpful comments on different drafts of this paper. This paper has been presented at the RCEA Time Series Econometrics Workshop in Madrid, at the 2026 IAAE in Lisbon, at the `5$^{th}$ Sailing the Macro Workshop', Siracusa September 2025, with the title: `\textit{Local fiscal multipliers in a data-scarce environment: The effectiveness of government spending across Italian regions}', at the `Macroeconometrics in Salerno Workshop', Salerno October 2025, with the title `\textit{Assessing the (in)effectiveness of government spending across Italian regions}', and at the `3$^{rd}$ Workshop in Sustainable Finance' of Spoke 4 GRINS, Venezia December 2025. The paper was written while Marco Mazzali was affiliated to the Department of Economics, University of Bologna. The study is part of the project GRIN P9 - Spoke 4 (WP-4) and was funded by the European Union - NextGenerationEU, in the framework of the “GRINS - Growing Resilient, INclusive and Sustainable project” (PNRR - M4C2 - I1.3 - PE00000018 – CUP J33C22002910001). The views and opinions expressed are solely those of the authors and do not necessarily reflect those of the European Union, nor can the European Union be held responsible for them. Giuseppe Cavaliere and Luca Fanelli also gratefully acknowledge financial support from the Italian Ministry of University and Research (PRIN 2022, Grant 20229PFAX5 002). 
            
			\noindent $^{a}$Department of Economics, University of Bologna, Italy.
			
            \noindent $^{b}$Department of Economics, University of Exeter Business School, UK. 

            \noindent $^{c}$Department of Economics and Finance, Università Cattolica del Sacro Cuore, Milan. 

		}
		\addtocounter{footnote}{-1}
		\renewcommand{\thefootnote}{\arabic{footnote}}%
		%EndExpansion
		{\normalsize \vspace{0.1cm} }
		
		{\large Giuseppe \textsc{Cavaliere}$^{a,b}$ , Luca \textsc{Fanelli}$^{a}$ , Marco \textsc{Mazzali}$^{c}$   }
		
		\medskip
		
	\end{center}

	\begin{abstract}  
	\singlespacing
       We propose a novel econometric methodology for Structural Vector Autoregressions with external instruments (`proxy-SVARs' or `SVAR-IVs') in panel data characterized by strong cross-sectional dependence, dynamic heterogeneity, and limited availability of direct external instruments for the shocks of interest. For each unit, we specify a Factor-Augmented proxy-SVAR (`proxy-FA-SVAR') that incorporates factors summarizing cross-sectional information from the non-policy variables of the system. The effects of the policy shocks are then recovered indirectly by estimating unit-specific policy reaction functions through a Minimum Distance approach. Identification relies on global instruments for the non-policy shocks; that is, proxies common to all units in the panel, internally constructed from a separate SVAR estimated on factors for the policy and non-policy variables. These global instruments can be complemented with local (idiosyncratic) instruments constructed from auxiliary unit-level SVARs. Their joint use renders the proxy-FA-SVARs overidentified and therefore statistically testable. We illustrate the methodology by estimating government spending multipliers for Italian NUTS-2 regions using annual data. The global and local instruments for the regional output shocks are obtained from Blanchard-Perotti-type SVARs.
       
	\medskip
     \noindent \textsc{Keywords}: Global instruments, Proxy-SVARs, Identification, Regional fiscal multipliers.
        
		\bigskip
	\end{abstract}

	\clearpage

\section{Introduction}

Empirical macroeconomists and policymakers are often interested in quantities such as multipliers, elasticities, and pass-through coefficients, which summarize how the economy responds dynamically to structural shocks. These objects are central to understanding the transmission of policy interventions to macroeconomic outcomes, and their accurate quantification is therefore of key importance. This task, however, becomes especially challenging when attention shifts from aggregate national economies to local settings such as regions, sectors, or other subnational units. By `local policy', we refer to policy interventions defined at the level of these subnational units, for example regional government spending, as opposed to aggregate national policy. Throughout the paper, we use the terms `region' and `unit' interchangeably to denote the cross-sectional units in the panel.

A prominent approach to identifying dynamic causal effects in empirical macroeconomics is the `proxy-SVAR' or `SVAR-IV' methodology. Proxy-SVARs rely on instruments, or proxies, to identify the structural shocks of interest; see \cite{stock2012disentangling,stock2018identification} and \cite{mertens2013dynamic}. In principle, this framework is flexible enough to be applied to regional or sectoral data, and more generally to settings in which the relevant phenomena have both a cross-sectional and a time-series dimension.

In practice, however, two major challenges arise. The first is cross-sectional heterogeneity and dependence. When local units differ in their dynamic responses to policy shocks, conventional approaches, including standard panel instrumental-variable methods, may fail to recover the causal effects of interest. For example, \cite{canova2024should} shows that the common practice of estimating fiscal multipliers in panel data is valid only under restrictive conditions, notably homogeneity in the dynamic responses of local units to policy changes. Relatedly, \cite{canova2025macroeconomic} provide evidence that the effects of EU regional structural funds differ markedly across European regions, reflecting variation in local institutions, fiscal capacity, and spending efficiency. These findings highlight the need for econometric frameworks that explicitly account for heterogeneity and interdependence across local units.

The second challenge is instrument scarcity. Credible external instruments for local policy shocks, such as exogenous innovations in regional fiscal spending, are often difficult to obtain because the relevant data are limited, unavailable, or measured imprecisely. This problem is particularly acute in applications where administrative and official statistics are typically less informative than at the national level; see, among others, \cite{ramey2016macroeconomic,nakamura2018identification,chodorow2019geographic}.

These issues are especially relevant for the estimation of fiscal multipliers in Italian regions, which motivates the methodological approach developed in this paper. More broadly, however, the proposed framework is not confined to this application. It is intended as a general econometric framework for identifying dynamic causal effects in panel data settings characterized by heterogeneous local responses, cross-sectional interdependence, and limited availability of valid external instruments for the shocks of interest.

We contribute to the literature by developing a novel econometric framework for local macroeconomic policy analysis that: (i) delivers region-specific dynamic causal effects while explicitly accounting for cross-sectional heterogeneity and dependence; (ii) remains applicable in environments where direct external instruments for local policy shocks are unavailable; and (iii) produces overidentified, and hence testable, models for structural policy evaluation. Methodologically, our approach builds on and extends the contributions of \cite{angelini2024identification} and \cite{canova2024should}. As in \cite{canova2024should}, we explicitly address cross-sectional heterogeneity through a region-wise approach. Unlike the existing literature, however, we recover the dynamic causal effects of interest through a methodology that combines three main elements.

The first is factor augmentation. We account for cross-unit dependence while preserving heterogeneity by incorporating common factors for the non-policy variables into each region-specific SVAR, thus obtaining Factor-Augmented SVARs (FA-SVARs) \cite[\`a la][]{bernanke2005measuring}. 

The second is an indirect estimation approach. To address the data-scarcity problem, namely, the limited availability of external instruments for the target (policy) shocks in official statistics and publicly available datasets, we extend the indirect identification strategy of \cite{angelini2024identification} to the case of proxy-FA-SVARs. Specifically, we estimate region-specific policy reaction functions using global instruments for the regional non-policy shocks; that is, proxies common to all units in the panel, which are extracted from the panel itself. We complement these global instruments with local (or idiosyncratic) ones, namely region-specific proxies, still extracted from the panel. Crucially, the joint use of global and local instruments renders the regional proxy-FA-SVARs overidentified, and hence empirically testable.

The third is instrument construction. The global and local instruments used to estimate the proxy-FA-SVARs are constructed, under the conditions detailed in the paper, from auxiliary SVARs estimated on variables that summarize the common and idiosyncratic components of the policy and non-policy variables in the panel. The identifying restrictions imposed on these auxiliary SVARs depend on the application at hand, but in general they must imply that the resulting instruments for the non-policy shocks are orthogonal to the policy shocks. Because of the data-scarcity constraint, these instruments are not drawn from external datasets but are constructed internally from the original panel. Their validity nevertheless rests on the same relevance and exogeneity conditions required in standard proxy-SVAR analysis. For expositional convenience, we retain the term `proxy-FA-SVAR' throughout the paper, while recognizing that the instruments used in these models are more precisely interpreted as internally constructed proxies for the non-policy shocks.

The proposed approach combines identifying assumptions imposed at different stages of the procedure. We discuss these assumptions and their implications in details in the methodological section of the paper.

To illustrate, suppose that only one non-policy variable and one policy variable are observed for each region. We then estimate region-specific policy reaction functions by Minimum Distance (MD), instrumenting the non-policy shock simultaneously with two proxies (the global and local one, respectively) and recovering the policy shock residually as the implied structural error term. This approach generalizes the logic developed in the fiscal literature by, among others, \cite{leeper1996does} and \cite{caldara2017analytics} and extended by \cite{angelini2024identification} to settings in which instruments for the policy shocks may be weak, while strong and exogenous instruments are available for the non-policy shocks. The global and local overidentifying instruments for the non-policy shock are constructed from separate auxiliary SVARs estimated on factors. In the empirical analysis of fiscal spending multipliers in Italian regions presented in Section \ref{sec:empirics}, for example, the global instrument for the regional output shock is obtained from a SVAR estimated on two factors extracted from the panel: one summarizing regional government spending and the other regional output. Following the recursive identification scheme of \cite{blanchard2002empirical}, we identify a common output shock that is, by construction, orthogonal to the common regional government spending shock, and use it as the global instrument. The local instrument is obtained analogously, from a separate SVAR estimated on the idiosyncratic counterparts of the same two factors.  It is worth noting that, in the fiscal literature, the use of shocks identified in an auxiliary step to aid structural identification is not without precedent; see, for example, \cite{ramey2018government}. Relative to \cite{ramey2018government}, however, we find it more convenient to construct instruments for the non-policy shocks of the system, and then use them to recover the policy shocks indirectly.

On the empirical side, as anticipated, the suggested methodology is applied to estimate fiscal multipliers for Italian regions at the NUTS-2 level, using annual data covering the period 1995--2021. Thus, the paper contributes to the recent empirical debate on the size of Italian regional fiscal multipliers \cite[see, e.g.,][]{deleidi2021quantifying, destefanis2022regional, lucidi2023misalignment}.
Relative to the existing literature, we find marked differences in fiscal policy effects across regions, highlighting the geographically uneven transmission of public spending. While the national multiplier suggests that Italian public spending is, on average, expansionary and quantitatively important, several regions, especially in the South and the Islands, struggle to translate additional fiscal resources into medium-term growth. Our bootstrap confidence intervals for the estimated regional multipliers, which explicitly account for potential nonlinear dependence in VAR disturbances, indicate that any assessment of the effectiveness of regional fiscal policy in Italy must confront the substantial uncertainty surrounding output responses to fiscal spending shocks. 
We also acknowledge explicitly that the limited time dimension of the available data, with only $T=27$ annual observations, calls for considerable caution in the interpretation of these empirical results. 

\bigskip 
\noindent\textit{Connections with the literature}
\medskip

This paper contributes to the econometric and empirical macroeconomic literature on the identification of dynamic causal effects in panel settings with both cross-sectional and time-series variation.  

Our framework shares with \cite{canova2024should}  the core insight that standard panel estimates may be severely biased when the data are generated by spatial macroeconomic processes, especially when units display heterogeneous dynamics. We explicitly address these concerns but, relative to \cite{canova2024should},  we formulate the analysis within a factor-augmented VAR framework with instruments. These instruments are constructed through a procedure designed to overcome the lack of data that can be used to instrument the shocks of interest. 

Our methodology also provides a transparent alternative, in applied work, to Bartik-type (shift-share) instruments. Bartik-type instruments are constructed by combining local shares with aggregate shifts \cite[see, e.g.,][]{bartik1991benefits, blanchard1992regional}. Despite their widespread use, their exogeneity cannot in general be taken for granted: consistency requires that initial local shares be uncorrelated with subsequent unobserved local factors, \cite[see][]{goldsmith2020bartik}. By contrast, in our framework exogeneity is achieved by design through structural identification in auxiliary SVARs estimated on suitably extracted factors from the panel. This procedure delivers a global instrument for the regional non-policy shock that is orthogonal to the regional policy shock. Moreover, whereas Bartik designs are typically implemented in cross-sectional or panel settings that often do not accommodate dynamic heterogeneity, our methodology explicitly allows for heterogeneous dynamics and cross-sectional dependence through region-specific proxy-FA-SVARs. 

Our approach also has points of contact with the granular instrumental variables (GIV) approach of \cite{gabaix2024granular}. In that framework, a suitably weighted sum of idiosyncratic shocks to a small number of granular units is used as an instrument for aggregate endogenous variables, for example to estimate demand and supply elasticities; see, among others, \cite{adrian2022term} and \cite{baumeister2024uncovering}. We also exploit the informational content of idiosyncratic shocks extracted from disaggregated data, but in a different way. Rather than aggregating them into a single national instrument for an aggregate equation, we construct unit-specific idiosyncratic instruments and combine them with a global instrument to obtain a vector of overidentifying proxies that capture two complementary dimensions of regional structural shocks. 

The paper is also related to \cite{bai2010instrumental}, who develop a factor-based methodology for constructing instrumental variables. Their key idea is that, if the common components extracted from a large panel capture the same sources of variation that drive the endogenous regressors, the estimated factors can be used directly as valid instruments. By contrast, our framework is designed for settings with instrument scarcity at the unit level, even when the cross-section remains reasonably large. We use factor modelling to summarize common co-movements across units and to exploit the information contained in the panel, but we do not use the factors directly as instruments. Instead, we use them to construct structurally identified global and local proxies, as discussed in Section \ref{sec:instruments}. 

Finally, our work engages directly with the literature on heterogeneous panels, sharing some of its core concerns while differing in both objectives and implementation. As in the Common Correlated Effects (CCE) approach of  \cite{pesaran2006estimation}, we model cross-sectional interdependence through common factors to avoid inconsistent estimation. Crucially, however, we use these factors not only as conditioning variables, but also as building blocks for constructing external instruments in environments with local instrument scarcity. Likewise, in the spirit of the Pooled Mean Group (PMG) estimator of  \cite{pesaran1999pooled}, we allow for dynamic heterogeneity across units. Unlike PMG, however, our framework does not impose long-run homogeneity restrictions. Its objective is the structural identification of local shocks through internally constructed instruments, and the estimation of their dynamic causal effects, rather than the estimation of common long-run parameters. Other prominent contributions to this literature include \cite{pesaran1995estimating}, \cite{im2003testing} and \cite{chudik2015common}, among many others. 

\bigskip 
\noindent\textit{Structure of the paper}
\medskip 

The remainder of the paper is organized as follows. Section \ref{sec:framework} presents the methodology, the proxy-FA-SVAR framework, the identification strategy, the construction of the global and local instruments, and a clarifying interpretation of the recovered policy shocks and of the salient aspects of the multi-step methodology adopted. Section \ref{sec:empirics} applies the methodology to estimate fiscal spending multipliers in Italian regions. Section \ref{sec:conclusion} concludes. Appendix A describes in detail how the common factors are extracted from the panel.

\section{Local policy evaluations through global and local instruments}\label{sec:framework}
	
This section presents the econometric methodology. Section \ref{sec:model_pt1} introduces the reference proxy-SVARs in a panel-data setting, focusing on the case in which instruments for the policy shocks are unavailable. Section \ref{sec:model_pt2} extends the analysis to region-specific proxy-FA-SVARs where policy shocks are identified indirectly using global and idiosyncratic instruments for the non-policy shocks. Section \ref{sec:instruments} details how the global and idiosyncratic instruments used to estimate and test the regional proxy-FA-SVARs are constructed. Section \ref{sec:interpretation} discusses the interpretation of the regional policy shocks identified by the proposed multi-step procedure, while Section \ref{sec:on-the-methodology}  highlights the main features and limitations of the framework, along with possible extensions.

\subsection{Regional proxy-SVARs under data scarcity}\label{sec:model_pt1}
	
Consider a panel of units $i = 1, \ldots, R$ and assume that, for each of these units, we observe $n$ variables, collected in the vector $Y_{i,t}:=(Y_{i,t}^{(1)},\ldots,Y_{i,t}^{(n)})^{\prime }$, observed for $t=1,\ldots,T$ time periods.  Throughout the paper, the cross-sectional dimension of the panel, $R$, is kept fixed, and we study estimators in the limit where the number of periods $T$ is treated as large. This setup allows us to identify and estimate $R$ distinct models for $Y_{i,t}$ separately, with cross-sectional dependencies (e.g., of the variables in $Y_{1,t}$ with those in $Y_{2,t}$) modelled as described below. We discuss in more detail in Section \ref{sec:interpretation} the relationship between the cross-sectional dimension, $R$, and the time series length, $T$, in our framework. 

We partition $Y_{i,t}$ as: $Y_{i,t}:=(Y_{i,t}^{p\prime }\text{ },Y_{i,t}^{o\prime})^{\prime }$, where $Y_{i,t}^{p}$ is $k\times 1$ and contains the subset of policy variables and $Y_{i,t}^{o}$ is $(n-k)\times 1$ and contains the outcome, or non-policy, variables of the system. In the empirical analysis of Section \ref{sec:empirics}, we consider $i=1,2,...,R=20$ Italian (NUTS-2 level) regions,  $Y_{i,t}^{p}$ contains real government spending in region $i$  ($k=1$), $Y_{i,t}^{o}$  real GDP in region $i$ ($n-k=1$), and $t=1,...,T=27$ are annual observations.  

For each region, $\mathcal{D}_{i,t}:=\sigma (Y_{i,t}\text{ },Y_{i,t-1},\ldots)$ is
the information set generated by the sequence ($Y_{i,t}\text{ },Y_{i,t-1},\ldots$), $\sigma (\cdot)$
being the sigma-field spanned by the variables in the argument. $\mathcal{D%
}_{t}^{j}:=\sigma
(Y_{1,t}^{(j)}\text{ },Y_{1,t-1}^{(j)}\text{ },\ldots;Y_{2,t}^{(j)}\text{ },Y_{2,t-1}^{(j)}\text{ },\ldots;Y_{R,t}^{(j)}\text{ },Y_{R,t-1}^{(j)},\ldots)
$ is the information set generated by the (single) $j$-th variable in $%
Y_{i,t}$ over time, for all regions. Similarly, $\mathcal{D}_{t}^{Y_{p}}$ and $%
\mathcal{D}_{t}^{Y_{o}}$ are the information sets generated by the policy and non-policy variables for all regions. Finally, $\mathcal{D}_{t}$ denotes the econometrician's information set at time $t$. It holds that $%
\cup _{i=1}^{R}\mathcal{D}_{i,t}\subseteq \mathcal{D}_{t}$ and $%
\cup _{j=1}^{n}\mathcal{D%
}_{t}^{j}\subseteq \mathcal{D}_{t}$.

For $i=1,\ldots,R$, it is assumed that $Y_{i,t}$ is generated by the reduced-form VAR($\ell$): 
\begin{equation}
	Y_{i,t} = \sum_{j = 1}^{\ell}\Pi_{i,j} Y_{i,t-j} + u_{i,t}\text{ }, \qquad t = 1, \ldots, T\label{eq:VAR}
\end{equation}
where $T$ is the number of time series observations (common to all regions), $ \Pi_i := (\Pi_{i,1}, \ldots, \Pi_{i,\ell}) $ is an $n \times n\ell$ matrix of autoregressive parameters and $ u_{i,t} $ is the vector of VAR disturbances with $ {E}(u_{i,t}) = 0 $ and ${E}(u_{i,t}u_{i,t}^{\prime}) =: \Omega_{i,u} $, $\ \Omega_{i,u} $ being a positive definite covariance matrix. For notational brevity, we do not include deterministic components in equation (\ref{eq:VAR}). Further regularity conditions on the VAR for $Y_{i,t}$ are provided in Section \ref{sec:model_pt2} below.
	
Let $\varepsilon_{i,t}$ be the $n$-dimensional vector of (latent) region-specific structural shocks. It is assumed that $\varepsilon_{i,t}$ is serially uncorrelated, i.e., ${E}(\varepsilon_{i,t}\varepsilon_{i,t-q}^{\prime}) = 0_{n\times n}$,  $q = \pm 1, \pm 2 \ldots$ , and uncorrelated across variables, i.e.,  ${E}(\varepsilon_{i,t}\varepsilon_{i,t}^{\prime}) = I_{n}$; the unit-variance normalization is made out of convenience and can be relaxed.

The relationship between the region-specific structural shocks and the region-specific VAR disturbances is given by 
	\begin{equation}
		u_{i,t} = B_i \varepsilon_{i,t}\text{ }, \qquad   \label{eq:mapping}
	\end{equation}
where $B_i$  is an $n \times n$  nonsingular matrix. The columns of $B_i$ capture the on-impact (instantaneous) effect of each structural shock in $\varepsilon_{i,t}$ on $Y_{i,t}$.  We partition the regional structural shocks as the variables $Y_{i,t}$, i.e., $ \varepsilon_{i,t} := (\varepsilon_{i,t}^{p\prime}\text{ }, \varepsilon_{i,t}^{o\prime})^{\prime}$, where the $k \times 1$ vector $\varepsilon_{i,t}^{p}$ collects the target structural shocks, also denoted policy shocks (e.g., fiscal policy shocks) and $\varepsilon_{i,t}^{o}$ contains the remaining $n-k$ non-target shocks (e.g., demand/supply shocks), also denoted non-policy shocks. Using this partition, system (\ref{eq:mapping}) specializes to the expression: 
	\begin{equation}
		u_{i,t} = B_{i, \scalebox{0.5}{$\bullet$}, p}\varepsilon_{i,t}^{p} + B_{i,\scalebox{0.5}{$\bullet$}, o}\varepsilon_{i,t}^{o} \label{eq:mapping_partition}
	\end{equation}
where $B_i := (B_{i,\scalebox{0.5}{$\bullet$}, p}\text{ }, B_{i,\scalebox{0.5}{$\bullet$}, o})$, and $B_{i,\scalebox{0.5}{$\bullet$}, p}$ and $B_{i,\scalebox{0.5}{$\bullet$}, o}$ are of dimensions  $n \times k$ and $n \times (n-k)$, respectively.

The Impulse Response Functions (IRFs), i.e., the responses of the variables in $Y_{i,t+h}$ to a one-standard-deviation change in the $j$-th policy shock in $\varepsilon_{i,t}^{p}$ are summarized in the expression: 
	\begin{equation}
		IRF_{i,j}(h) := (S_n(C_{i,\Pi})^hS_n^{\prime})B_{i,\scalebox{0.5}{$\bullet$},p}e_j, \quad 1 \leq j \leq k \label{eq:IRF}
	\end{equation}
where $S_n := (I_n, 0_{n \times n(\ell-1) })$ is a selection matrix and $e_j$ is a selection vector containing all zeros except in the $j$-th position containing `$1$', which therefore selects the $j$-th column of $B_{i,\scalebox{0.5}{$\bullet$}, p}$.  The ultimate object of the analysis is the estimation of relative (normalized) dynamic responses of the outcome variables at time $t+h$, $Y_{i,t+h}^o$, $h=0,1,\ldots$, to policy shocks in $\varepsilon_{i,t}^{p}$ of assigned magnitudes.\footnote{The IRFs in (\ref{eq:IRF}) are absolute responses. Relative responses are a suitable transformation of the absolute responses in (\ref{eq:IRF}).}   

Conventional proxy-SVAR estimation hinges on the existence of a vector $z_{t}$ of $r$ instruments ($r\geq k$ ) in $\mathcal{D}_{t}$ which are (strongly) correlated with $\varepsilon_{i,t}^{p}$ (relevance condition), and are uncorrelated with $\varepsilon_{i,t}^{o}$ (exogeneity condition). In principle, these instruments can also be region-specific, meaning that it might be the case that $z_{t}\in \mathcal{D}_{i,t}$. When a vector of instruments $z_{t}$ exists and is relevant and exogenous, it can be used to estimate the IRFs in (\ref{eq:IRF}), along the lines suggested in, e.g., \cite{mertens2013dynamic} and \cite{stock2018identification}.
In this paper, we address the scenario where, because of data scarcity, it is not possible to implement the conventional proxy-SVAR approach.  Specifically, for any region $i$, there does not exist a vector of $r\geq k$ external instruments $z_{t}\in \mathcal{D}_{t}$, such that $z_{t}$ is relevant for the policy shocks $\varepsilon_{i,t}^{p}$ and exogenous for the non-policy shocks $\varepsilon_{i,t}^{o}$. While this setup does not exclude the possibility that, in general, the econometrician's information set might contain valid instruments for the non-policy shocks (i.e., variables correlated  with $\varepsilon_{i,t}^{o}$ and uncorrelated with $\varepsilon_{i,t}^{p}$), a key purpose of this paper is to provide a general methodology that delivers instruments for the regional non-policy shocks. These instruments are obtained from the estimation of auxiliary SVARs estimated on factors extracted from $\mathcal{D}_{t}^{Y_{o}}$.  This implies that, strictly speaking, the proxies we use for the non-policy shocks are not `external' to the original panel of data but, rather, are generated internally. Instrument validity, nevertheless, rests on the usual relevance and exogeneity conditions required in standard proxy-SVAR analysis.

While fully capturing regional heterogeneity, a region-wise approach solely based on the proxy-SVARs  in (\ref{eq:VAR})-(\ref{eq:mapping_partition}) omits, by construction, the possible dynamic interrelations characterizing variables in different regions. For example, in the empirical illustration presented in Section \ref{sec:empirics}, it is reasonable to expect that, due to geographical, historical, and economic reasons, the dynamics of GDP  in the `Lombardia' region is interrelated with the dynamics of GDP in the `Piemonte' region. Obviously, not accounting for this regional dependence is akin to omitting relevant regressors in linear regression models, which would affect estimator consistency. In the next section, we address this issue by augmenting the baseline region-specific proxy-VARs with factors extracted from the set $\mathcal{D}_{t}^{Y_{o}}$,  intended to capture cross-unit dependence.         

Driven by these insights, the methodology we design in the next section combines the proxy-SVAR framework introduced above with a multi-step procedure which requires:
\begin{enumerate}
    \item Augmenting each regional proxy-VAR for $Y_{i,t}$ with a vector of factors $F_t^{o}$ extracted from the set $\mathcal{D}_{t}^{Y_o}$ ($F_t^{o} \in \mathcal{D}_{t}^{Y_o}$), intended to capture cross-regional dependencies through dynamic interrelations characterizing the non-policy variables. This results in $R$ regional proxy-FA-SVARs.
    \item Instrumenting the non-policy shocks of each regional proxy-FA-SVAR, mirroring the approaches in \cite{caldara2017analytics} and \cite{angelini2024identification}. We use a set of overidentifying instruments for the non-policy shocks which combine global and local variables; global instruments are proxies common to all $R$ regional models; local instruments are specific to the $i$-th region in the panel. Therefore, the  moment conditions characterizing the $R$ regional proxy-FA-SVARs are testable. 
    \item Building the global and local instruments for the non-policy shocks from separate SVAR models estimated on factors extracted from the panel. These SVAR models are identified with the purpose of generating relevant and exogenous proxies for the regional non-policy shocks.  
\end{enumerate}

In the next sections, we provide the specific details characterizing this multi-step methodology and clarify the nature of the dynamic causal effects identified with this approach for each region.

\subsection{Regional proxy-FA-SVARs and the indirect identification strategy}\label{sec:model_pt2}

In this section, we augment the regional proxy-SVARs with factors intended to capture cross-regional dependencies among the non-policy variables. Subsequently, we describe our indirect identification strategy, consisting in the estimation of region-specific dynamic causal effects by instrumenting the regional non-policy shocks.

To account for cross-sectional dependencies through non-policy variables, we extract a vector of factors, say $F_t^{o}$, from the set of non-policy variables of the panel ($F_t^{o} \in \mathcal{D}_{t}^{Y_o}$).  Incidentally, this process also delivers vectors of region-specific, idiosyncratic factors for the non-policy variables, say $\eta_{i,t}^{o} $ , $i=1,\ldots,R$, which will be used in Section \ref{sec:instruments} to build local instruments for the non-policy shocks. Appendix A briefly illustrates how the non-policy global factor $F_t^{o}$, and the non-policy idiosyncratic factors $\eta_{i,t}^{o}$ can be obtained from $\mathcal{D}_{t}^{Y_o}$.  

\bigskip 
\noindent\textit{Regional proxy-FA-SVARs and policy reaction functions}
\medskip

For each region, we consider the $m$-dimensional `augmented' vector (dimensions are reported alongside blocks):
\begin{equation}
    W_{i,t} := 
    \left(
    \begin{array}{c}
         Y_{i,t}  \\
         F_{t}^o
    \end{array}
    \right)
    \quad
    \begin{array}{c}
         n \times 1  \\
         (n-k) \times 1 
    \end{array}
    \label{eq:augmented_variables}
\end{equation}
where $F_t^{o}$ is $ (n-k) \times 1 $ and $ m = n + (n-k)$. The dimension of the vector $F_t^{o}$ is discussed below. We assume that the data-generating process for $W_{i,t}$ belongs to the VAR system:
\begin{equation}
    W_{i,t} = \sum_{j = 1}^{\ell} \Gamma_{i,j} W_{i,t-j} + w_{i,t}  ,\quad w_{i,t} := 
    \left(
    \begin{array}{c}
         u_{i,t}  \\
         f_{i,t}^{o} 
    \end{array}
    \right) , 
    \quad
    \begin{array}{cc}
         i = 1, \ldots, R; \\
          t = 1, \ldots, T.
    \end{array}
    \label{eq:FAVAR}
\end{equation}
In model (\ref{eq:FAVAR}), $\Gamma_{i,j}$ are $m\times m$ matrices of autoregressive coefficients and $w_{i,t}$ is a vector of residuals with zero mean and variance ${E}(w_{i,t}w_{i,t}^{\prime}) = \Omega_{i,w}$. $f_{i,t}^{o}$ denotes the disturbance term associated with the marginal VAR equations for the factors, and $u_{i,t}:=(u_{i,t}^{p\prime}\text{ }, u_{i,t}^{o\prime})^{\prime}$ is partitioned as $Y_{i,t}:=(Y_{i,t}^{p\prime  }\text{ },Y_{i,t}^{o\prime})^{\prime}$. Relative to the VAR (\ref{eq:VAR}), the augmented system (\ref{eq:FAVAR}) features the dynamic interactions possibly characterizing region-specific (policy and non-policy) variables and the factors driving all non-policy variables in the panel, while still accounting for regional heterogeneity through the parameters in the matrices $\Gamma_{i,j}$ and $\Omega_{i,w}$, respectively. 

The vector of factors ${F}^o_t$ in (\ref{eq:augmented_variables})-(\ref{eq:FAVAR})  is assumed to be $ (n-k) \times 1 $, meaning that each element of ${F}^o_t$ can be interpreted as a weighted average, across regions, of each non-policy variable in the system. In principle, in the spirit of FA-SVAR models \`a la \cite{bernanke2005measuring}, ${F}^o_t$ can be replaced with a sub-vector $\tilde{F}^o_t$ of lower dimension, $\dim(\tilde{F}^o_t)<(n-k)$, implying that the dimension of $W_{i,t}$ can possibly be smaller than $m$. More parsimonious specifications are appealing in scenarios where $(n-k)$ is considerably larger relative to $k$. However, as it will be clear in Section \ref{sec:instruments}, we do need to rely on the $(n-k)$-dimensional vector of factors ${F}^o_t$ in the construction of global instruments for the non-policy shocks. With this in mind, and to simplify notation, in the rest of the paper we continue using ${F}^o_t$ in the augmented vector $W_{i,t}$.

The regularity conditions we place on model (\ref{eq:FAVAR}) are those typical of proxy-SVAR analysis, and are summarized in the assumption that follows. We denote by $C_{i,\Gamma}$ the companion matrix associated with the FA-VAR model (\ref{eq:FAVAR}).  

\begin{assumption}
    For each region $i$, the observations $W_{i,1}$,\ldots,$W_{i,T}$ are generated by the FA-VAR in (\ref{eq:FAVAR}), where the companion matrix $C_{i,\Gamma}$ is stable (meaning that all eigenvalues of $C_{i,\Gamma}$ lie inside the unit disk), and the disturbances $w_{i,t}:=(u_{i,t}^{\prime}\text{ },f_{i,t}^{o\prime})^{\prime }$ satisfy the following conditions: \newline
    (i) the process $\left\{ w_{i,t}\right\} $ is a strictly stationary weak white noise with $m\times m$ positive definite, unconditional covariance matrix ${{E}}(w_{i,t}w_{i,t}^{\prime }):=\Omega _{i,w}<\infty $; \newline
    (ii) the process $\left\{ w_{i,t}\right\} $ satisfies the \textit{alpha}-mixing conditions in Assumption 2.1 of \cite{bruggemann2016inference}; \newline
    (iii) $w_{i,t}$ has absolutely summable cumulants up to order eight.
    \label{assumption:2}
\end{assumption}

According to Assumption \ref{assumption:2}, the enlarged VAR system (\ref{eq:FAVAR}) is asymptotically stable. This rules out scenarios in which the variables in $W_{i,t}$, and hence, both $Y_{i,t}$ and the factors ${F}^o_t$, are generated by I(1) processes. This is consistent either with the case in which the original panel variables have been transformed to ensure stationarity, or with the possibility that the VAR roots, while stable, generate highly persistent dynamics. Assumption \ref{assumption:2}.(i)-(ii) further specifies that the disturbance term $w_{i,t}$ forms an \textit{alpha}-mixing process, thus covering a broad class of uncorrelated but potentially nonlinearly dependent processes, including disturbances with conditional heteroskedasticity of unknown form. This covers a wide range of empirically relevant scenarios. Assumption \ref{assumption:2}.(iii) ensures consistency of the moving block bootstrap (MBB) \cite[][]{jentsch2022asymptotically}. The MBB is therefore the recommended method for constructing asymptotically valid confidence bands under Assumption \ref{assumption:2}.

By pre-multiplying system (\ref{eq:FAVAR}) by the $m \times m$ nonsingular matrix $ A_i$, we obtain the regional SVAR:
\begin{equation}
    A_iW_{i,t} = \sum_{j = 1}^{\ell} \Upsilon_{i,j} W_{i,t-j} + v_{i,t}\text{ }, 
\end{equation}
where $\Upsilon_{i,j} := A_{i}\Gamma_{i,j}$, $j=1,\ldots,\ell$ and $  A_iw_{i,t}=v_{i,t}$. In this structural model, the (sub)system of equations in $ A_iw_{i,t}=v_{i,t}$  for which the regional policy shocks are on the right-hand side is given by:
\begin{equation}
		A_{i,p,\scalebox{0.5}{$\bullet$}} w_{i,t} = A_{i,p,p} u_{i,t}^{p} + A_{i,p,o} u_{i,t}^{o}+A_{i,p,f} f_{i,t}^{o}  = \varepsilon_{i,t}^{p} \label{eq:Amapping}
	\end{equation}
where $A_{i,p,\scalebox{0.5}{$\bullet$}} := (A_{i,p,p}, A_{i,p,o} , A_{i,p,f})$ is a $k \times m$ matrix (of full row rank $k<m$) collecting the first $k$ rows of $A_i$, and $A_{i,p,p}$, $A_{i,p,o}$ and $A_{i,p,f}$  are defined accordingly. Under proper identifying restrictions, it is in principle possible to recover the regional policy shocks $\varepsilon_{i,t}^{p}$ from system (\ref{eq:Amapping}), then tracking the dynamic causal effects of these shocks on the variables $Y_{i,t+h}$, $h=0,1,\ldots$  

Specifically, we notice that by exploiting the nonsingularity of the $k \times k$ matrix $A_{i,p,p} $ and imposing the zero restrictions $A_{i,p,f}:=0_{k\times (n-k)}$ which prevent the factors from entering region $i$'s structural equation, system (\ref{eq:Amapping}) can be rewritten in the form:
\begin{equation}
		u_{i,t}^{p} = K_{i,o} u_{i,t}^{o}  + K_{i,p}\varepsilon_{i,t}^{p} \label{eq:policy_function}
\end{equation}
Equation (\ref{eq:policy_function}) can be interpreted as a system of policy reaction functions in which the regional policy variables (adjusted for the VAR dynamics) respond to the regional non-policy variables (adjusted for the VAR dynamics) and are shifted by the regional policy shocks; $K_{i,o}:= -A_{i,p,p}^{-1}A_{i,p,o}$ is $k \times (n-k)$, and is interpretable as the elasticity of region $i$'s policy variables with respect to region $i$'s non-policy variables; $K_{i,p}:=  A_{i,p,p}^{-1}$ is $k \times k$ nonsingular, and reads as a scaling factor associated with the regional policy shocks. It is worth remarking that the restrictions $A_{i,p,f}:=0_{k\times (n-k)}$ only characterize the structural specification, not the reduced-form system's dynamics where the non-policy factors are allowed to interact with region $i$'s variables.

The structure of the policy reaction functions (\ref{eq:policy_function}) suggests that using a vector of instruments correlated with region $i$'s non-policy shocks, $\varepsilon_{i,t}^{o}$, and uncorrelated with region $i$'s policy shocks, $\varepsilon_{i,t}^{p}$, the parameters can be estimated consistently. Below we show that the policy reaction functions can be estimated by a MD approach. Suppose now we have the MD estimates of $A_{i,p,p} $ and $A_{i,p,o} $ (under $A_{i,p,f}:=0_{k\times (n-k)}$), denoted $\hat{A}_{i,p,p} $ and $\hat{A}_{i,p,o}$, respectively. Then, $\hat{K}_{i,o}:=-\hat{A}_{i,p,p}^{-1}\hat{A}_{i,p,o}$ and $\hat{K}_{i,p}:=\hat{A}_{i,p,p}^{-1}$, implying that replacing the VAR disturbances with the reduced-form residuals, the regional policy shocks are estimated as: 
\begin{equation*}
\hat{\varepsilon}_{i,t}^{p} = \hat{K}_{i,p}^{-1}( \hat{u}_{i,t}^{p} - \hat{K}_{i,o} \hat{u}_{i,t}^{o}) = \hat{A}_{i,p,p}\hat{u}_{i,t}^{p}+\hat{A}_{i,p,o}\hat{u}_{i,t}^{o} , \qquad  t = 1, \ldots, T
\end{equation*}
Accordingly, for each regional proxy-FA-SVAR whose empirical validity is not empirically rejected (see below), the IRFs in (\ref{eq:IRF}) can be estimated as follows: $C_{i,\Pi}$ is replaced with the corresponding VAR-based OLS reduced-form estimates, while the instantaneous coefficients in $B_{i,\scalebox{0.5}{$\bullet$} ,p}$ are recovered by using the relationship: 

\begin{equation*}
B_{i,\scalebox{0.5}{$\bullet$} ,p}=\Sigma _{i,u}\left( 
\begin{array}{c}
A_{i,p,p}^{\prime } \\ 
A_{i,p,o}^{\prime }%
\end{array}%
\right) 
\end{equation*}
which characterizes proxy-SVAR analysis \cite[][]{angelini2024identification}.

\bigskip 
\noindent\textit{Minimum distance estimation and inference}
\medskip

The variables we use to instrument $u_{i,t}^{o} $ in the estimation of the policy reaction functions (\ref{eq:policy_function}) region-wise are collected in the vector $ g_{i,t} := ( g_{t}^{\prime}\text{ }, \bar{g}_{i,t}^{\prime})^{\prime}$, where $g_{t}$ contains $n-k$ global instruments for $\varepsilon_{i,t}^{o}$,  i.e., common to all $R$ regions in the panel, and $\bar{g}_{i,t}$  contains $n-k$ local (or idiosyncratic) instruments for $\varepsilon_{i,t}^{o}$, i.e., specific to region $i$.  We detail in Section \ref{sec:instruments} how the instruments in $g_{i,t}$ are built in practice. We impose that $g_{i,t}$ satisfies the following conditions:
\begin{eqnarray}
E(g_{i,t}\varepsilon _{i,t}^{o\prime}) &=&\Psi _{o} \text{ }, \quad rank[\Psi
_{o}]=n-k  \label{eq:relevance_non_policy} \\
E(g_{i,t}\varepsilon _{i,t}^{p\prime}) &=&0_{2(n-k)\times k} \qquad \qquad \qquad \qquad i=1,\ldots,R  \label{eq:exogeneity_policy}
\end{eqnarray}
namely, that $g_{i,t}$ is correlated with the regional non-policy shocks and is uncorrelated with the regional policy shocks. Conceptually, the role of the instruments $g_{i,t}$ is to approximate the non-policy shocks in each region via two complementary channels: (i) a global source of information, which aims to capture the common dimension behind the realization of regional non-policy shocks; (ii) a local source of information, which is intended to pick up the idiosyncratic (sectoral or regional) aspects characterizing these shocks. Moreover, as will be shown below, instruments $g_{i,t}$ that jointly satisfy the relevance (\ref{eq:relevance_non_policy}) and the exogeneity conditions (\ref{eq:exogeneity_policy}) overidentify the regional non-policy shocks. 

The MD estimation of the policy reaction functions (\ref{eq:policy_function}) via the instruments $g_{i,t}:=(g_{t}^{\prime }\text{ },\bar{g}_{i,t}^{\prime })^{\prime }$ mimics the approach put forth in \cite{angelini2024identification}. In particular, taking the variance of both sides of system (\ref{eq:Amapping}) we obtain the moment conditions:
\begin{equation}
A_{i,p,\scalebox{0.5}{$\bullet$} }\Omega _{i,w}A_{i,p,\scalebox{0.5}{$\bullet$} }^{\prime }=I_{k}  \label{m1}
\end{equation}%
while multiplying system (\ref{eq:Amapping}) on the right-hand side by $g_{i,t}^{\prime }$, taking expectations and exploiting the exogeneity condition (\ref{eq:exogeneity_policy}), we obtain the instrument-based moment conditions:%
\begin{equation}
A_{i,p,\scalebox{0.5}{$\bullet$} }\Omega _{i,w,g}=0_{k\times 2(n-k)}  \label{m2}
\end{equation}%
where $\Omega _{i,w,g}:=E(w_{i,t}g_{i,t}^{\prime })$ is $m \times 2(n-k)$, and the parameters in $A_{i,p,\scalebox{0.5}{$\bullet$} }$ are restricted such that: $A_{i,p,f}=0_{k\times (n-k)}$. 

Jointly considered, (\ref{m1})-(\ref{m2}) imply $(1/2)k(k+1) +2k(n-k)$ moment conditions and $k^2+k(n-k)$ free parameters in the matrices $A_{i,p,p}$ and $A_{i,p,o}$, respectively. Therefore, if also the necessary and sufficient rank condition in Proposition 1 in \cite{angelini2024identification} holds, the MD estimation of the regional policy reaction functions by the instruments $g_{i,t}$ can be carried out by replacing the reduced-form parameters in $\Omega _{i,w}$ and $\Omega _{i,w,g}$ in (\ref{m1})-(\ref{m2}) with their consistent estimators, then minimizing the resulting distance function. This gives, for each estimated regional proxy-FA-SVAR, $k(n-k)-(1/2)k(k-1)$ testable overidentifying restrictions.

Let $\alpha _{i}$ denote the $d$-dimensional ($d=k^{2}+k(n-k)$) parameter vector containing the free, unrestricted elements in the matrices $A_{i,p,p}$ and $A_{i,p,o}$, respectively. Let $\alpha _{i,0}$ be the vector of corresponding true parameters. Furthermore, we denote by $\hat{Q}_{T}(\alpha _{i})$
the MD criterion function obtained by summarizing the moment conditions  (\ref{m1})-(\ref{m2}) and replacing the reduced-form parameters 
$\Omega _{i,w}$ and $\Omega _{i,w,g}$ by their consistent estimators $%
\hat{\Omega}_{i,w}:=\frac{1}{T}\sum_{t=1}^{T}\hat{w}_{i,t}\hat{w}%
_{i,t}^{\prime }$ and $\hat{\Omega}_{i,w,g}:=\frac{1}{T}\sum_{t=1}^{T}\hat{%
w}_{i,t}g_{i,t}^{\prime }$, respectively. 

The following proposition provides the regularity conditions under which asymptotic inference in the regional proxy-FA-SVARs is standard.

\begin{proposition}
 Let $\hat{\alpha}_{i,T}$ be the MD estimator resulting from the minimization of $\hat{Q}_{T}(\alpha _{i})$. If under
Assumption \ref{assumption:2}, the following conditions hold: the instruments $g_{i,t}:=(g_{t}^{\prime }\text{ },\bar{g}_{i,t}^{\prime })^{\prime }$ satisfy (\ref{eq:relevance_non_policy}) and (\ref{eq:exogeneity_policy}); the parametric restrictions $A_{i,p,f}=0_{k\times (n-k)}$ hold; the criterion function $\hat{Q}_{T}(\alpha _{i})$ satisfies the regularity conditions that make $\alpha _{i}$ identifiable locally; then
\begin{equation}
    \hat{\alpha}_{i,T}\overset{p}{\rightarrow }\alpha _{i,0}     ; \qquad \sqrt{T}(\hat{\alpha}_{i,T}-\alpha _{i,0})\overset{d}{\rightarrow } N(0_{d},V_{\alpha _{i}}). \label{eq:asymptotics}
\end{equation}
\begin{equation}
T\hat{Q}_{T}(\hat{\alpha}_{i,T})\overset{d}{\rightarrow }\chi_{\nu}^2 \label{eq:test}
\end{equation} 
where $\nu = k(n-k)-(1/2)k(k-1)$ is the number of overidentifying restrictions.
\label{proposition:1}
\end{proposition}

The proof of Proposition \ref{proposition:1} follows from a straightforward adaptation of Propositions 1 and 2 in \cite{angelini2024identification}. Note that as regards inference on the parameters $\alpha _{i}$, the asymptotic covariance matrix ${V}_{\alpha _{i}}$ can be estimated by the MBB. 

Proposition \ref{proposition:1} also ensures that if the overidentifying instruments used for the regional non-policy shocks are relevant and exogenous, the regional proxy-FA-SVAR estimated by the MD approach can be tested empirically. This can be done by computing the test statistic in (\ref{eq:test}), whose asymptotic distribution is standard under the null and diverges either when the instruments $g_{i,t}:=(g_{t}^{\prime }\text{ },\bar{g}_{i,t}^{\prime })^{\prime }$ do not satisfy the exogeneity condition (\ref{eq:exogeneity_policy}), or when the restrictions $A_{i,p,f}=0_{k\times (n-k)}$ on the policy reaction functions are not valid. 

The next section explains how the instruments $g_{i,t}:=(g_{t}^{\prime }\text{ },\bar{g}_{i,t}^{\prime })^{\prime }$ for the regional non-policy shocks are obtained.

\subsection{Global and idiosyncratic instruments}\label{sec:instruments}

In this section, we focus on the construction of the instruments $g_{i,t} := (g_t^{\prime}\text{ },\bar g_{i,t}^{\prime})^{\prime}$, the key ingredients of the identification and estimation strategy of the proxy-FA-SVAR models presented in the previous section.

Given the $(n-k)\times 1$ vector of factors $F^o_t$ introduced in Section \ref{sec:model_pt1}, we consider the analogue $ k\times 1$ vector of policy factors, denoted $F^p_t$; see Appendix A for details. The joint $n$-dimensional vector $F_t=(F^{p\prime}_t\text{ },F^{o\prime}_t)^{\prime}$  summarizes, for each (policy and non-policy) variable at time $t$, the information across all regions. Thus, $F_t=(F^{p\prime}_t\text{ },F^{o\prime}_t)^{\prime}$ can be viewed as the counterpart of the region-specific variables $Y_{i,t}:=(Y_{i,t}^{p\prime }\text{ },Y_{i,t}^{o\prime
})^{\prime }$, effectively reflecting the aggregate, common information characterizing all $R$ units in the panel.    
Similarly, for each region $i$, we collect the idiosyncratic factors (see Appendix A) in the $n$-dimensional vector $\eta_{i,t} := (\eta^{p\prime}_{i,t}\text{ }, \eta^{o\prime}_{i,t})^{\prime}$, interpretable as region-specific counterparts of the global factors $F_t=(F^{p\prime}_t\text{ }, F^{o\prime}_t)^{\prime}$ . 

We extract the vector of global instruments $g_{t}$ from a SVAR specified and estimated on the variables $F_t=(F^{p\prime}_t\text{ },F^{o\prime}_t)^{\prime}$ , and the $R$ vectors of local instruments $\bar g_{i,t}$ from the specification and estimation of $R$ separate SVARs for the variables $\eta_{i,t}=(\eta^{p\prime}_{i,t}\text{ },\eta^{o\prime}_{i,t})^{\prime}$, respectively.  

\bigskip 
\noindent\textit{SVAR for the global factors}
\nobreak\par\nopagebreak
\medskip

We approximate the law of motion of the factors $F_t$ by the VAR($l$):

\begin{equation}
F_{t}=\sum_{j=1}^{l}\Theta _{j}F_{t-j}+e_{t}\text{ },\text{ \ \ }e_{t}=\left( 
\begin{array}{c}
e_{t}^{p} \\ 
e_{t}^{o}%
\end{array}%
\right) 
\label{eq:F_VAR}
\end{equation}
where $e_{t}$ is the associated disturbance term, partitioned for simplicity as $F_t$.

The SVAR counterpart is obtained by pre-multiplying both sides of system (\ref{eq:F_VAR}) by the nonsingular matrix:
\begin{equation*}
G:=\left( 
\begin{array}{cc}
G_{p,p} & G_{p,o} \\ 
G_{o,p} & G_{o,o}%
\end{array}%
\right) 
\label{eq:G_partitions}
\end{equation*}
where $G_{p,p}$, $G_{p,o}$, $G_{o,p}$ and $G_{o,o}$ are blocks of conformable dimensions. This yields the SVAR model:

\begin{equation}
GF_t = \sum_{j=1}^l\Theta^G_j F_{t-j} + \xi_t,
\label{eq:F_SVAR}
\end{equation}
where $\Theta^G_j=G\Theta _{j}$, $j=1,\ldots,l$, and the mapping between reduced-form disturbances and structural shocks can be summarized by the equations:
\begin{equation}
Ge_{t}=\left( 
\begin{array}{c}
G_{p,p}e_{t}^{p}+G_{p,o}e_{t}^{o} \\ 
G_{o,p}e_{t}^{p}+G_{o,o}e_{t}^{o}%
\end{array}%
\right) =\left( 
\begin{array}{c}
\xi _{t}^{p} \\ 
\xi _{t}^{o}%
\end{array}%
\right) 
\label{eq:F_mapping}
\end{equation}

Under proper identification restrictions on $G$, we can interpret the elements in $\xi_t=(\xi_t^{p\prime}\text{ }, \xi_t^{o \prime})^{\prime}$ as global structural shocks, i.e., common to all regions. Our ultimate objective is to label the elements in $\xi _{t}^{p}$  as global policy shocks and the elements in $\xi _{t}^{o}$ as global non-policy shocks, respectively.

The structural shocks $\xi_t=(\xi_t^{p\prime}\text{ }, \xi_t^{o \prime})^{\prime}$ are such that:%

\begin{equation*}
E\left( \left( 
\begin{array}{c}
\xi _{t}^{p} \\ 
\xi _{t}^{o}%
\end{array}%
\right) (\xi _{t}^{p\prime}\text{ },\xi _{t}^{o\prime})\right)
=\left( 
\begin{array}{cc}
I _{k} & 0_{k\times (n-k)} \\ 
0_{(n-k)\times k} & I_{n-k}%
\end{array}%
\right). 
\end{equation*}

The global instrument for the regional non-policy shocks can be recovered from the SVAR (\ref{eq:F_SVAR})-(\ref{eq:F_mapping}) by placing at least $(1/2)n(n-1)$ identification restrictions on the matrix $G$, estimating the model (if also a proper necessary and sufficient rank condition holds) by standard methods, and then defining:
\begin{equation}
g_{t}:=\hat{\xi}_{t}^{o}:=\hat{G}_{o,p}\hat{e}_{t}^{p} + \hat{G}_{o,o}\hat{e}_{t}^{o}, \qquad t=1,\ldots,T;
\label{eq:global_instrument}
\end{equation}                            
where $\hat{G}_{o,p}$ and $\hat{G}_{o,o}$ here denote the estimates of the identified parameters in $G_{o,p}$ and $G_{o,o}$, respectively. 

The identification restrictions on the matrix $G$ required for constructing the instrument in (\ref{eq:global_instrument}) depend on the specific phenomenon at hand. For example, in the empirical illustration we present in Section \ref{sec:empirics} in the estimation of regional government spending multipliers, $F^{p}_t=F^{gov}_t$ is a factor summarizing information on regional government spending and $F^{o}_t=F^{gdp}_t$ is a factor summarizing information on regional output ($k=1; n=2$). Following \cite{blanchard2002empirical}, we specify the matrix $G$ to be lower triangular:
\begin{equation*}
G:=\left( 
\begin{array}{cc}
G_{p,p} & 0 \\ 
G_{o,p} & G_{o,o}%
\end{array}%
\right) 
\end{equation*}
implying that government spending does not respond to the output shock on impact but only with lags.

\bigskip
\noindent\textit{SVARs for the local factors}
\nobreak\par\nopagebreak
\medskip

To build the local instruments $\bar g_{i,t}$, for each region $i$ we estimate a SVAR model for the idiosyncratic factors $\eta_{i,t} = (\eta_{i,t}^{p\prime}\text{ },\eta_{i,t}^{o\prime})^{\prime}$ , whose identification mirrors the model considered in (\ref{eq:F_SVAR}).  Specifically, each SVAR is given by: 
\begin{equation}
G_i\eta_{i,t} = \sum_{j=1}^{l_i} \Theta^G_{i,j} \eta_{i,t-j} + {\xi}_{i,t}
 \text{ \ \ , \ \ }
  G_{i}e_{i,t}=\xi_{i,t} 
 \text{ \ \ , \ \ }
 i = 1,\ldots,R
\label{eq:eta_SVAR}
\end{equation}
where $ l_{i}$ is the number of VAR lags, ${G}_i$ is the idiosyncratic analogue of $G$ and is partitioned as in (\ref{eq:F_SVAR}), $\Theta^G_{i,j} := G_{i}\Theta_{i,j}$ is the analogue of $\Theta^G_{j}$ and $e_{i,t} := ({e}^{p\prime}_{i,t}\text{ },{e}^{o \prime}_{i,t})^{\prime}$ are reduced-form innovations with covariance matrix $\Omega_{i,e}$; finally,  ${\xi}_{i,t} := ({\xi}^{p\prime}_{i,t}\text{ },{\xi}^{o \prime}_{i,t})^{\prime}$ are the region-specific structural shocks. In this case, the objective is to label the elements in ${\xi}^{p}_{i,t}$ as local policy shocks and the elements in ${\xi}^{o}_{i,t}$ as local non-policy shocks, respectively.

We identify the local structural shocks in \eqref{eq:eta_SVAR} by adapting the identification strategy used for the global factors, unless there are specific economic reasons for imposing region-specific restrictions. Therefore, in general, the identification restrictions on the matrices ${G}_i$ follow from those placed on $G$. 

The local instruments for the regional non-policy shocks are then recovered as:
\begin{equation}
    \bar g_{i,t} := \hat{\xi}^{o}_{i,t} = \hat{G}_{i,o,p} \hat{e}_{i,t}^{p} + \hat{G}_{i,o,o} \hat{e}_{i,t}^{o} , \qquad t=1,\ldots,T;
\end{equation}
where, again, $\hat{G}_{i,o,p}$ and $\hat{G}_{i,o,o}$ are the estimates of the identified parameters in $G_{i,o,p}$ and $G_{i,o,o}$, respectively. 

For each region, $g_{i,t} := (g_t^{\prime}\text{ },\bar g_{i,t}^{\prime})^{\prime}$ is used as the vector of overidentifying instruments in the estimation of the proxy-FA-SVAR models.

\subsection{Interpretation of the policy shocks and multipliers}\label{sec:interpretation}

This section discusses the interpretation of the regional policy shocks and the corresponding regional dynamic causal effects recovered through the approach described above.

Policy shocks are recovered from the regional policy reaction functions in equation~\eqref{eq:policy_function}, estimated by instrumenting the non-policy shocks with a vector of overidentifying global and local instruments. The estimated policy shock for region $i$,  $\hat{\varepsilon}_{i,t}^{p}$, should therefore be interpreted as the component of the reduced-form innovation in regional government spending that cannot be explained by: (i) the region-specific non-policy innovation identified through the proxy structure; and (ii) the lagged regional and common dynamics embedded in the factor-augmented SVAR. In this sense, the procedure isolates a policy innovation specific to region $i$, conditional on the common macroeconomic environment summarized by the factors $F^o_t$. If these identifying conditions fail, however, the recovered policy shock may capture not only the intended policy innovation, but also measurement error, omitted non-policy disturbances, or misspecification of the regional policy reaction function. To the extent that these failures invalidate the overidentifying restrictions, they may be detected by the data through the corresponding specification tests. 

The so-identified regional policy shocks are not meant to coincide with aggregate national fiscal shocks, nor should they be interpreted as purely idiosyncratic disturbances in a narrow sense. Rather, they are local structural innovations recovered from region-specific policy equations and identified conditional on common factors that absorb pervasive co-movements in non-policy variables across regions. To the extent that aggregate fiscal programs, such as nationwide or EU-funded spending plans, are implemented with heterogeneous timing, composition, or local absorption capacity, their regional incidence may be reflected in these shocks. By contrast, the suggested approach is not designed to estimate the effects of a single aggregate fiscal intervention that is homogeneous across all regions. Similarly, cross-regional spillovers are accommodated only insofar as they are captured by the common factors entering the regional systems, rather than being separately identified through bilateral or network-specific transmission channels. We turn to the role of regional spillovers in the concluding section of the paper. 

This distinction is important for interpreting the Italian regional fiscal multipliers estimated in Section \ref{sec:empirics}.  In this section, the estimated impulse responses and implied cumulative multipliers measure the dynamic effect of an exogenous one-unit innovation in regional government spending, as identified by our procedure, on regional output. Hence, they should be read as local multipliers associated with region-specific spending innovations after controlling for common macroeconomic conditions, rather than as multipliers for a fully aggregate national program. At the same time, these local multipliers remain informative for policy episodes such as the Italian National Recovery and Resilience Plan (NRRP), since large aggregate programs are ultimately implemented through region-specific spending realizations whose effects may differ systematically across territories. Our estimates are therefore most naturally interpreted as measuring the heterogeneous local transmission of fiscal interventions within a common aggregate policy environment.

\subsection{On the multi-step approach}
\label{sec:on-the-methodology}

Before moving to the empirical part, it is useful to clarify some features and implications of the multi-step nature of our methodology.

As remarked earlier, the regional dynamic causal effects are recovered in three steps. First, the policy and non-policy factors are extracted from the panel (see Appendix A for details). Second, the global and local instruments $g_{i,t}=(g_t',\bar g_{i,t}')'$ are constructed from the auxiliary SVARs discussed in Section~\ref{sec:instruments}. This intermediate step is required because direct external instruments for the policy shocks are unavailable for the estimation of the regional proxy-FA-SVARs in the third step. The identification strategy therefore proceeds indirectly, by generating instruments for the non-policy shocks and then using them to recover the regional policy shocks. In this sense, the regional proxy-FA-SVARs rely on internally constructed proxy instruments, rather than on direct external instruments observed outside the panel. Third, the regional proxy-FA-SVARs are estimated by MD using these instruments, as discussed in Section~\ref{sec:model_pt2}. Confidence intervals for the quantities of interest, namely the regional fiscal multipliers estimated in Section~\ref{sec:empirics}, are based on the proxy-FA-SVARs estimated in this final step.

The factors play a distinctive role in this procedure. In the regional proxy-FA-SVARs, $F^o_t$ enters as a vector of additional regressors intended to capture cross-regional interdependence that would otherwise be omitted from the model. This differs from standard FA-SVAR settings, where factors summarize information from large datasets containing heterogeneous variables, such as prices, quantities, surveys, and financial indicators. In our panel, by contrast, for each variable $j$, the collection $\{Y^{(j)}_{i,t}\}_{i=1}^R$ contains regional realizations of the same underlying quantity. The factors $F_t^{o} \in \mathcal{D}_{t}^{Y_o}$ can be viewed as weighted cross-regional averages of the outcome variables. Moreover, because $R$ is fixed by design, the factor extraction step does not rely on a large-cross-section asymptotic argument. We therefore do not require large-$R$ factor consistency, nor conditions such as $\sqrt{T}/R \to 0$, which are typically invoked to justify treating estimated factors as observed regressors \cite[][]{baing2006}. Accordingly, our interpretation of the factors does not hinge on a strong-factor assumption in the usual approximate-factor sense.  Consistent estimation of the regional proxy-FA-SVARs under Assumption \ref{assumption:2} instead relies on $T \to \infty$.
In the empirical illustration presented in the next section, $T=27$ (annual observations from 1995 to 2021) is only slightly larger than $R=20$ (the number of Italian NUTS-2 regions), so the finite-sample approximation to the underlying asymptotic framework should be regarded with due caution. This limitation does not invalidate the proposed methodology, but it does call for prudence in the interpretation of estimated fiscal multipliers.

At the same time, $F^o_t$ is a generated regressor, and so is the vector of instruments $g_{i,t}$; both are used in the same sample from which they are constructed. We quantify uncertainty by applying the MBB only in the final step of the procedure.\footnote{Bootstrap methods for factor-augmented models have been studied, among others, by \cite{goncalves2014}, \cite{goncalves2020}, and \cite{yamamoto2019}. Their validity typically relies on a double-asymptotic regime in which both $R$ and $T$ diverge at suitable rates, a framework that is difficult to reconcile with the empirical set-up we face in Section \ref{sec:empirics}.} The resulting confidence intervals therefore do not explicitly account for the uncertainty arising from factor extraction and instrument construction in the earlier steps. The MBB confidence intervals reported for the Italian regional fiscal multipliers in Section~\ref{sec:empirics} should thus be interpreted with this limitation in mind, and may understate true sampling uncertainty.
In principle, the proposed methodology could be extended to incorporate all layers of uncertainty implied by the multi-step procedure, not just the uncertainty in the last step. A fully integrated bootstrap would require, for each bootstrap replication of the panel: (i) re-extracting the factors, thereby obtaining a bootstrap analogue of $F^o_t$; (ii) re-estimating the auxiliary SVARs to obtain bootstrap analogues of the instruments $g_t$ and $\bar g_{i,t}$; and (iii) re-estimating the regional proxy-FA-SVARs and the implied dynamic causal effects. We leave the development of such a procedure, and the formal proof of its consistency, to future work.

A final remark concerns the apparent tension, in our framework, between recursive identification and proxy-based identification. In the second step of the procedure, auxiliary SVAR models are estimated on factors and idiosyncratic components to construct global and local instruments for the non-policy shocks, which are then used in the regional proxy-FA-SVARs. Identification in these auxiliary SVARs can be achieved by standard methods, including recursive schemes. In the fiscal application of Section~\ref{sec:empirics}, for example, we rely on Blanchard-Perotti-type specifications, which order government spending before output.

Recursive SVARs and proxy-FA-SVARs therefore play distinct roles in our framework. In this sense, recursive identifying assumptions are not eliminated, but relocated to the auxiliary systems used to construct the instruments, while identification in the regional models remains proxy-based and empirically testable through overidentifying restrictions. The former are used only to construct overidentifying instruments for the non-target shocks, given the scarcity of instruments for the target shocks, while the latter use these instruments to recover the regional policy shocks. This distinction is important. Under a purely recursive identification scheme imposed directly on the regional models, the specification would not be overidentified and therefore would not allow for the same form of empirical testing. By contrast, a distinctive feature of the proposed proxy-FA-SVAR approach is that the instruments are deliberately constructed to overidentify the regional models. As a result, each regional specification is subjected to empirical scrutiny through overidentifying restrictions.

\section{Regional fiscal multipliers in Italy}\label{sec:empirics}

The estimation of regional fiscal multipliers in Italy provides a natural empirical application of the proposed framework. On the one hand, the Italian case is characterized by substantial cross-regional heterogeneity and strong spatial interdependence. On the other hand, the scarcity of regional fiscal instruments makes it difficult to apply standard proxy-SVAR methods directly. These features make Italian regions a suitable environment in which to illustrate the empirical relevance of our methodology.

This section is organized as follows. Section \ref{sec:empirics_data} describes the data.  Section \ref{sec:empirics_instruments} discusses the construction and strength of the global and local instruments. Section \ref{sec:empirics_multipliers} defines the fiscal multipliers. Section \ref{sec:empirics_results} presents the empirical results. 

\subsection{Data}\label{sec:empirics_data}

Europe is administratively divided into territorial units, classified according to the Nomenclature of Territorial Units for Statistics (NUTS). This classification system comprises several hierarchical layers: NUTS-0 represents the national level (e.g., Italy); NUTS-1 corresponds to macro areas within the country; and NUTS-2 corresponds to the regional level within the country.

Our analysis is primarily based on Italian NUTS-2 units, as these are the territorial levels typically considered in the empirical literature on local fiscal policy. According to the classification established in 2016, this encompasses 20 administrative regions (where the autonomous provinces of Bolzano and Trento are included as part of the Trentino-Alto Adige region).

We also consider the NUTS-1 classification, which groups these regions into five macro areas: North-West (Piemonte, Valle d'Aosta, Liguria, and Lombardia), North-East (Trentino-Alto Adige, Veneto, Friuli-Venezia Giulia, and Emilia-Romagna), Centre (Toscana, Umbria, Marche, and Lazio), South (Abruzzo, Basilicata, Calabria, Campania, Molise, and Puglia), and Islands (Sardegna and Sicilia). When convenient for discussion, we refer to the South and Islands macro areas jointly as the  Mezzogiorno. 

Time series data are sourced from the Italian Institute of Statistics (ISTAT) and are freely downloadable from the \textquotedblleft\emph{Conti e aggregati economici territoriali}' (Territorial Accounts and Economic Aggregates) section of the national accounts. For each NUTS-2 unit, our annual time series span the period 1995--2021.\footnote{As such, our analysis can only partially evaluate the impact of increased fiscal spending in the aftermath of the COVID-19 pandemic on regional economies.}

We focus on two key regional economic variables: Gross Domestic Product ($GDP$) and Total Government Expenditure ($Gov$).  $Gov$ is obtained as the sum of Government Consumption and Investment. All variables are expressed in real per-capita terms, deflated using the national GDP deflator (computed as the ratio between national GDP at current prices and national GDP at constant prices). Annual average population at the regional level is provided by ISTAT. Unless explicitly indicated, all variables are expressed in logarithmic units.  

Our empirical analysis is based on the vector of variables $ Y_{i,t} = (gov_{i,t}, gdp_{i,t})^{\prime}$, where $gov_{i,t}$ is logged real per-capita $Gov$ and $gdp_{i,t}$ is logged real per-capita $GDP$. Hence, $n=2$ and $k=1$, where the policy block is given by $Y_{i,t}^{p} = gov_{i,t}$ and the non-policy (outcome) block by $Y_{i,t}^{o} = gdp_{i,t}$. The cross-sectional unit index, $i$, runs across $R=20$ regions, associated with the NUTS-2 classification detailed above. The time index $t$ runs from 1995 to 2021, for a total of $T=27$ annual observations. 

In line with the developments in Section \ref{sec:model_pt2}, cross-regional dependencies are captured via the vector of non-policy factors $F_{t}^o$, which in this setup collapses to a scalar denoted by $F_t^{gdp}$. This factor is obtained as the first principal component of the regional GDP series, and is constructed while explicitly accounting for the possible non-stationarity of the regional GDP time series (\citealp[]{barigozzi2021large}; see Appendix A for details). It should be noted that, in principle, $F_t^{gdp}$ can be derived from various aggregation levels immediately preceding the level characterizing the analysis.  For instance, when dealing with NUTS-2 regions, $F_t^{gdp}$ could potentially be extracted from all regions in the panel or, alternatively, solely from the regions in the NUTS-1 macro area to which region $i$ belongs, or from adjacent macro areas. We do not observe substantial differences in empirical results from measuring $F_t^{gdp}$ in these different ways. 

The augmented vector in (\ref{eq:augmented_variables}) is given by:
\begin{equation*}
    W_{i,t} = 
    \left(
    \begin{array}{c}
    gov_{i,t}\\
    gdp_{i,t}\\
    F_{t}^{gdp}
    \end{array}
    \right).
\end{equation*}
The regional proxy-FA-SVARs for $W_{i,t}$ are estimated using data classified at the NUTS-2 level alone. Results referring to the NUTS-1 and NUTS-0 levels are obtained by suitably aggregating the estimated fiscal multipliers as detailed below.

\subsection{Instruments}\label{sec:empirics_instruments}

To build the global and local instruments for the regional output shocks, we follow the strategy outlined in Section \ref{sec:instruments}. We extract the global factors $F_t^o = F_t^{gdp}$,  $F_t^p = F_t^{gov}$ and the idiosyncratic factors $\eta_{i,t}^{o}=\eta_{i,t}^{gdp}$, $\eta_{i,t}^{p}=\eta_{i,t}^{gov}$, $i=1,\ldots,R$ from the panel, and  estimate the $R+1$ SVARs as discussed in Section \ref{sec:instruments}. These models are identified using a Blanchard-Perotti-type recursive scheme where government spending is assumed not to respond on impact to the output shock. 

\begin{figure}[h!]
    \centering
    \makebox[\textwidth][c]{\includegraphics[width=1.2\textwidth]{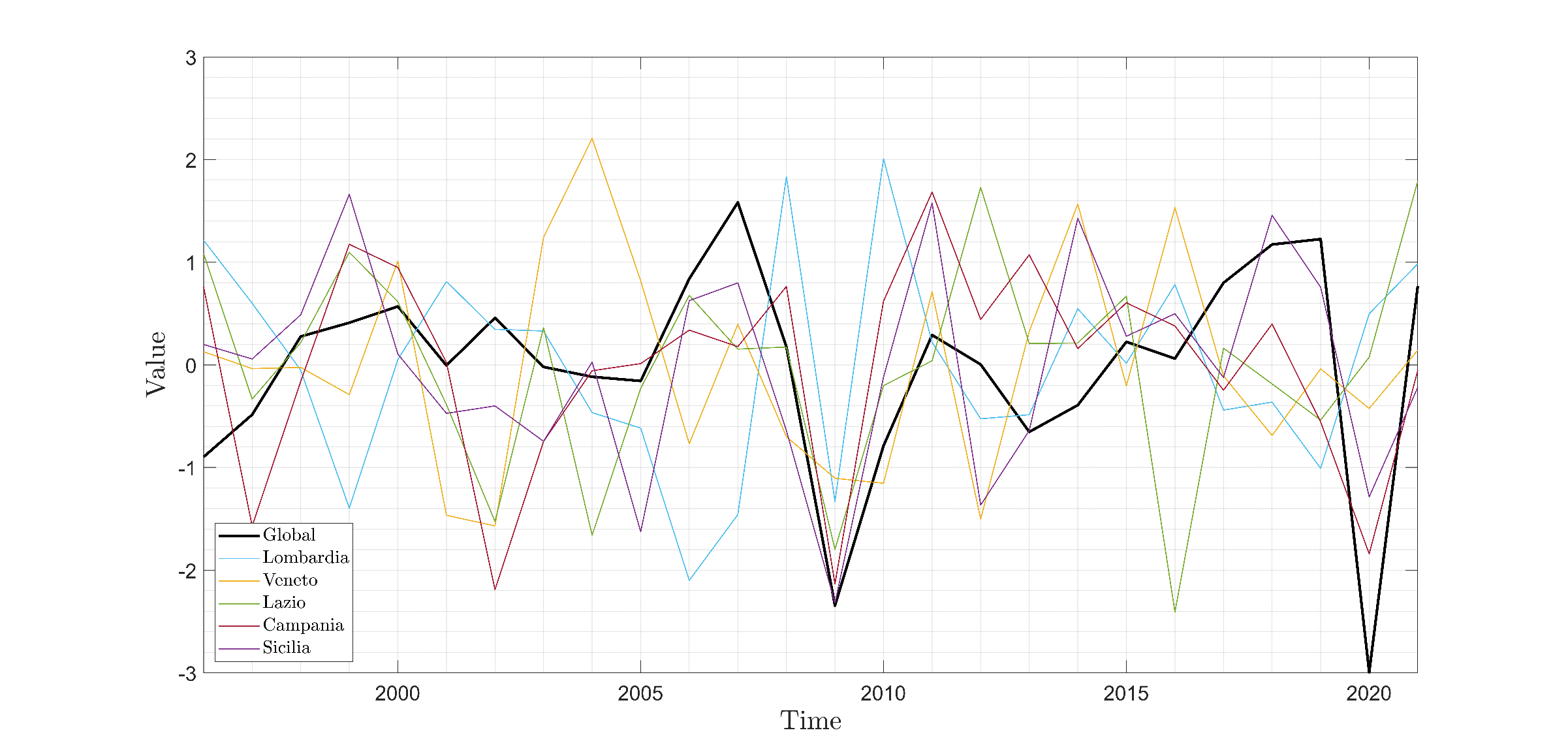}}
    \caption{Global vs (selected) local instruments}
    \label{fig:instruments_plot}
\end{figure}

This approach provides, for each region, the vector of instruments $g_{i,t} = (g_t'\text{ }, \bar{g}_{i,t}')'$ which overidentify the government spending reaction functions featured by the proxy-FA-SVAR models. The implied fiscal spending multipliers are discussed in Section \ref{sec:empirics_results}.

Figure~\ref{fig:instruments_plot}  compares the global instrument $g_t$ with a selection of regional (local) instruments $\bar{g}_{i,t}$ for Lombardia, Veneto, Lazio, Campania, and Sicilia. All series capture major macroeconomic episodes, including the 2009 crisis, but the figure also shows that the local instruments contain variation not absorbed by the global one. This impression is consistent with the correlation structure: the correlations between the global instrument and the selected local instruments are moderate, while pairwise correlations among the local instruments are generally smaller and heterogeneous. Taken together, these patterns suggest that $g_t$ primarily captures the common component of regional output shocks, whereas the local instruments capture idiosyncratic regional variation.

As in standard proxy-SVAR analysis, we assess instrument strength through first-stage regressions. Following   \cite{lewis2025robust}, we report the heteroskedasticity-robust effective $F$-statistic of \cite{olea2013robust}. Table \ref{tab:first_stage_reg} shows that, for all regions, the statistic exceeds the conventional threshold value of $23.1$. This provides evidence against weak-instrument concerns and supports the relevance of the constructed instrument vectors in the estimation of the regional policy reaction functions. 

\begin{table}[h!]
    \caption{Heteroskedasticity-robust first-stage F-tests}
    \centering
    \small
    \renewcommand{\arraystretch}{1.1} 
    \begin{tabular}{l@{\hspace{4cm}}r} 
\midrule 
Region & $F$-stat  \\
\midrule 
\multicolumn{1}{l}{                      Piemonte}   &   31.46    \\
\multicolumn{1}{l}{                 Valle d'Aosta}   &  142.82    \\
\multicolumn{1}{l}{                       Liguria}   &   35.66    \\
\multicolumn{1}{l}{                     Lombardia}   &   47.30    \\
\multicolumn{1}{l}{                        Veneto}   &   31.49    \\
\multicolumn{1}{l}{         Friuli-Venezia Giulia}   &   68.47    \\
\multicolumn{1}{l}{                Emilia-Romagna}   &   61.74    \\
\multicolumn{1}{l}{           Trentino-Alto Adige}   &   41.11    \\
\multicolumn{1}{l}{                       Toscana}   &   49.13    \\
\multicolumn{1}{l}{                        Umbria}   &  107.76    \\
\multicolumn{1}{l}{                        Marche}   &   63.87    \\
\multicolumn{1}{l}{                         Lazio}   &   26.36    \\
\multicolumn{1}{l}{                       Abruzzo}   &   44.04    \\
\multicolumn{1}{l}{                        Molise}   &   69.76    \\
\multicolumn{1}{l}{                      Campania}   &   33.52    \\
\multicolumn{1}{l}{                        Puglia}   &   34.01    \\
\multicolumn{1}{l}{                    Basilicata}   &   57.04    \\
\multicolumn{1}{l}{                      Calabria}   &   31.31    \\
\multicolumn{1}{l}{                       Sicilia}   &   35.61    \\
\multicolumn{1}{l}{                      Sardegna}   &   29.40    \\
\midrule 
\end{tabular}
    
    \subcaption*{
    \footnotesize 
    \renewcommand{\baselineskip}{11pt}
    \textbf{Note:} For each region, the table reports \citeauthor{olea2013robust}'s \citeyearpar{olea2013robust} effective $F$-statistic for the regression of $\hat{u}_{i,t}^{gdp}$ on the instrument vector $g_{i,t}$.}
    \label{tab:first_stage_reg}
\end{table}

\subsection{Fiscal multipliers}\label{sec:empirics_multipliers}
	
We define the fiscal multiplier as the euro response of output to an effective one-euro change in the fiscal variable. Following \cite{mountford2009effects}, we compute the cumulative multiplier at horizon $h$ (where $h$ indicates the number of periods after the government spending shock occurs) as the ratio of the cumulative response of output over the entire $h$-period horizon to the cumulative response of the fiscal variable over the same period; formally:
	\begin{equation}
		\mathcal{M}_{i,gov}(h) = \frac{\sum_{j=0}^{h}IRF_{i,gdp,gov}(j)}{\sum_{j=0}^{h}IRF_{i,gov,gov}(j)} \times c_{i,gov}  \label{eq:MU_mult}
	\end{equation}
where $c_{i,gov}$ is a region-specific scaling factor that converts elasticities into the euro equivalent; see below. Letting $GDP$ be the (unlogged) level of output and $Gov$ be the (unlogged) level of total government expenditure, we set the scaling factor to the sample average of the ratio $GDP/Gov$; see \textit{inter alia} \cite{caldara2017analytics} and \cite{angelini2024identification}. Moreover, since the scalars $c_{i,gov}$ may differ significantly across regions, we compute them using their macro area level (NUTS-1) counterparts \cite[][]{lucidi2023misalignment}.\footnote{This definition of the fiscal multipliers enhances the comparability of our results with those documented in the relevant literature. Alternatively, one may consider variables in levels divided by a measure of potential output, as for instance in \cite{gordon2010end}, \cite{bernardini2020heterogeneous} and \cite{ramey2018government}. See \cite{ramey2019ten} for a discussion.} 

The macro area (NUTS-1) and national (NUTS-0) multipliers are obtained by aggregating regional estimates at the NUTS-2 level. Specifically, NUTS-1 multipliers are computed as weighted averages of the regional multipliers within each macro area, using their respective GDP share as weights. Similarly, NUTS-0 multipliers are computed by aggregating the regional multipliers still using GDP weights.

\subsection{Empirical results}\label{sec:empirics_results}

Table \ref{tab:regional_multipliers} reports the estimated regional fiscal multipliers for Italian regions over the period 1995--2021. For each region, we report the impact multiplier $\mathcal{M}_{g}(0)$, which captures the instantaneous (at horizon $h=0$) response of output to a fiscal shock, the peak multiplier $ \mathcal{M}_{g}(peak)$, which captures the peak value of the multiplier over the horizon considered; and the long-run multiplier $\mathcal{M}_{g}(\infty)$, approximated by $\mathcal{M}_{g}(h)$ at horizon $h=10$, together with  $68\%$ MBB confidence intervals computed on 1000 replications. The last column reports the bootstrap $p$-value of the overidentifying-restrictions test.

The results reveal considerable heterogeneity across regions, underscoring that the effectiveness of government spending is profoundly shaped by local economic structures and conditions. We notice that none of the regional proxy-FA-SVARs is rejected at conventional significance levels, so the data do not provide evidence against the maintained identifying moment conditions. At the same time, the confidence intervals show that the uncertainty surrounding estimated multipliers is substantial.

\begin{table}[h!]
    \caption{Italian regional fiscal multipliers} 
    \renewcommand{\arraystretch}{1.3} 
    \centering
    \resizebox{\textwidth}{!}{
    \begin{tabular}{l cc cc cc c} 
\midrule 
Regions & $\mathcal{M}_{g}(0)$ & $68\%\text{-}CI$ & $\mathcal{M}_{g}(peak)$ & $68\%\text{-}CI$ & $\mathcal{M}_{g}(\infty)$ & $68\%\text{-}CI$ & \shortstack[c]{\small Bootstrap\\ \small p-value} \\
\midrule
Piemonte&  3.550 & [ 0.723,  3.565]   &    3.550( 0) & [ 0.723,  3.565]    &    2.164 & [ 1.409,  2.277] &    0.51 \\
Valle d'Aosta&  0.954 & [-0.305,  0.946]   &    1.988(10) & [ 0.189,  2.245]    &    1.988 & [ 0.189,  2.245] &    0.78 \\
Liguria&  2.624 & [ 0.787,  2.972]   &    3.758(10) & [ 2.070,  4.071]    &    3.758 & [ 2.070,  4.071] &    0.73 \\
Lombardia&  3.550 & [ 2.033,  3.759]   &    3.550( 0) & [ 2.033,  3.759]    &    2.873 & [ 2.258,  3.013] &    0.32 \\
Veneto&  3.241 & [ 1.228,  3.337]   &    3.241( 0) & [ 1.228,  3.337]    &    2.152 & [ 1.732,  2.632] &    0.69 \\
Friuli-Venezia Giulia&  1.649 & [-0.147,  1.929]   &    2.088( 5) & [ 0.874,  2.274]    &    2.068 & [ 0.984,  2.292] &    0.55 \\
Emilia-Romagna&  2.734 & [ 1.419,  3.827]   &    3.471( 6) & [ 3.316,  3.570]    &    3.463 & [ 3.417,  3.490] &    0.49 \\
Trentino-Alto Adige&  0.785 & [-0.316,  2.459]   &    0.983(10) & [-1.144,  4.890]    &    0.983 & [-1.144,  4.890] &    0.23 \\
Toscana&  3.042 & [ 2.144,  3.351]   &    3.042( 0) & [ 2.144,  3.351]    &    2.342 & [ 1.875,  2.448] &    0.15 \\
Umbria&  2.067 & [ 0.981,  2.328]   &    2.405( 3) & [ 1.758,  2.699]    &    2.344 & [ 1.842,  2.597] &    0.42 \\
Marche&  2.020 & [ 0.942,  2.436]   &    2.549( 9) & [ 1.808,  3.044]    &    2.549 & [ 1.810,  3.050] &    0.38 \\
Lazio&  2.729 & [ 1.847,  3.223]   &    2.729( 0) & [ 1.847,  3.223]    &    2.200 & [ 2.221,  3.960] &    0.69 \\
Abruzzo&  0.969 & [ 0.169,  1.017]   &    0.980( 3) & [ 0.095,  1.156]    &    0.965 & [-0.238,  1.335] &    0.76 \\
Molise&  1.054 & [ 0.650,  1.146]   &    1.054( 0) & [ 0.650,  1.146]    &    0.828 & [ 0.537,  1.238] &    0.79 \\
Campania&  1.738 & [ 1.350,  2.111]   &    1.738( 0) & [ 1.350,  2.111]    &    1.440 & [ 1.334,  1.625] &    0.63 \\
Puglia&  1.601 & [ 1.041,  1.866]   &    1.664( 2) & [ 1.325,  1.848]    &    1.576 & [ 1.295,  1.799] &    1.00 \\
Basilicata&  0.920 & [ 0.196,  0.951]   &    0.920( 0) & [ 0.196,  0.951]    &    0.626 & [-0.139,  0.728] &    0.72 \\
Calabria&  0.881 & [-1.121,  1.651]   &    1.601(10) & [-1.525,  2.189]    &    1.601 & [-1.525,  2.189] &    0.53 \\
Sicilia&  1.360 & [ 0.965,  1.539]   &    1.360( 0) & [ 0.965,  1.539]    &    1.334 & [ 1.150,  1.405] &    0.44 \\
Sardegna&  1.234 & [ 0.950,  1.468]   &    1.234( 0) & [ 0.950,  1.468]    &    1.233 & [ 0.947,  1.437] &    0.45 \\
\midrule
\end{tabular}

    }

    \subcaption*{
    \footnotesize 
    \renewcommand{\baselineskip}{11pt}
    \textbf{Note:} The table displays: (i) impact multipliers, $\mathcal{M}_{g}(0)$; (ii) peak  multipliers, $ \mathcal{M}_{g}(peak)$ with the corresponding peak horizon in brackets; and (iii) the long-run multipliers  $\mathcal{M}_{g}(\infty)$, approximated at the horizon $h=10$. Point estimates are accompanied by $68\%$ MBB confidence intervals. The last column reports, for each region, the bootstrap p-values associated with the overidentifying restriction tests computed on 1000 MBB replications.}
    \label{tab:regional_multipliers}
\end{table}

The estimates reveal a marked North--South gradient. Northern and, to a lesser extent, central regions typically display larger and more persistent multipliers. Piemonte, Lombardia, and Veneto stand out, with impact multipliers around 3.5 and long-run multipliers above 2. In these regions, the peak tends to occur on impact, suggesting a relatively rapid transmission of spending shocks to economic activity. Liguria, Emilia-Romagna, Toscana, Umbria, Marche, and Lazio also display sizeable multipliers, generally somewhat smaller than those in the North-West but still economically important. In some of these regions, the peak response occurs only after a few years, indicating a more gradual propagation of the fiscal shock.

By contrast, southern regions and the Islands exhibit smaller multipliers on average. Campania records the largest impact multiplier within the Mezzogiorno, but still well below the levels observed in the North. In other southern and Island regions, such as Molise, Puglia, Sardegna, and Sicilia, impact multipliers are generally above one, but remain comparatively modest. For Abruzzo and Basilicata, the estimated multipliers remain below one throughout the horizon and long-run effects are imprecisely estimated. Calabria and Trentino-Alto Adige are the two cases in which confidence intervals remain wide at all horizons, so that the data are less informative about the magnitude and even the sign of the regional responses.

\begin{table}[h!]
    \caption{Italian macro areas and national fiscal multipliers}
    \renewcommand{\arraystretch}{1.3} 
    \centering
    \resizebox{\textwidth}{!}{
    \begin{tabular}{l cc cc cc} 
\midrule 
Regions & $\mathcal{M}_{g}(0)$ & $68\%\text{-}CI$ & $\mathcal{M}_{g}(peak)$ & $68\%\text{-}CI$ & $\mathcal{M}_{g}(\infty)$ & $68\%\text{-}CI$  \\
\midrule
Italy&  2.625 & [ 1.448,  2.641]   &    2.625( 0) & [ 1.448,  2.641]    &    2.285 & [ 1.882,  2.667] \\
North-West&  3.445 & [ 1.597,  3.377]   &    3.445( 0) & [ 1.597,  3.377]    &    2.772 & [ 2.103,  2.846] \\
North-East&  2.625 & [ 1.241,  3.087]   &    2.799( 2) & [ 2.172,  3.195]    &    2.531 & [ 2.026,  3.055] \\
Centre&  2.703 & [ 1.924,  2.927]   &    2.703( 0) & [ 1.924,  2.927]    &    2.291 & [ 2.130,  3.192] \\
South&  1.444 & [ 0.957,  1.595]   &    1.444( 0) & [ 0.957,  1.595]    &    1.388 & [ 0.763,  1.529] \\
Islands&  1.326 & [ 0.966,  1.466]   &    1.326( 0) & [ 0.966,  1.466]    &    1.307 & [ 1.111,  1.385] \\
\midrule
\end{tabular}

    }
    \subcaption*{
    \footnotesize 
    \renewcommand{\baselineskip}{11pt}
\textbf{Note:} The table displays: (i) impact multipliers, $\mathcal{M}_{g}(0)$; (ii) peak  multipliers, $ \mathcal{M}_{g}(peak)$, along with the corresponding peak horizon in brackets; and (iii) the long-run multipliers  $\mathcal{M}_{g}(\infty)$, approximated at the horizon $h=10$. The macro area multipliers are computed as weighted averages of the regional multipliers within each area, using their respective GDP share as weights, while the national multipliers are obtained by GDP-weighting all regional estimates. Each point estimate is accompanied by the $68\%$ MBB confidence intervals.}
    \label{tab:aggregate_multipliers}
\end{table}

Table \ref{tab:aggregate_multipliers} aggregates the regional estimates to the NUTS-1 and NUTS-0 levels. The same North--South gradient emerges clearly. The North-West displays the largest multipliers, followed by the North-East and the Centre, whereas the South and the Islands show markedly smaller responses. At the national level, the implied government spending multiplier is positive and economically sizeable, at about 2.6 on impact and 2.3 in the long run.

Overall, the empirical evidence points to substantial geographical heterogeneity in the transmission of fiscal policy across Italian regions. The results are broadly in line with the existing evidence for Italy, which also documents larger multipliers in the Centre-North than in the South and the Islands. Relative to the recent contributions in  \cite{deleidi2021quantifying}, \cite{destefanis2022regional}, \cite{lucidi2023misalignment} and \cite{matarrese2023italian}, our results are obtained within a framework that explicitly allows for regional heterogeneity, cross-sectional dependence, and overidentification-based specification testing.

\section{Concluding remarks and possible extensions}\label{sec:conclusion}

We have proposed a novel econometric methodology that extends the proxy-SVAR framework to a panel data setting characterized by two key challenges. The first is instrument scarcity, namely the difficulty of finding valid external instruments for the policy shocks of interest when data availability is limited. The second is heterogeneity and interdependence, that is, a setting in which the dynamic causal effects of interest may differ across units while substantial cross-unit linkages and spillovers are present.

In our framework, the cross-sectional dimension of the panel is treated as fixed, while the time dimension is allowed to be large. Admittedly, in the empirical application considered in this paper, where we estimate regional fiscal multipliers for Italy, the time dimension is only moderately larger than the cross-sectional dimension, with $T=27$ annual observations and $R=20$ NUTS-2 regions. We do not view this relatively limited time dimension as an obstacle to the application of our methodology. While the moderate time dimension naturally calls for some caution in the interpretation of the empirical results, it does not undermine the usefulness of the proposed approach. Indeed, one of the main motivations of our framework is precisely to provide a coherent way of conducting structural analysis in environments where the available information is scarce but, overall, the cross-sectional structure of the panel is economically informative and can be conveniently exploited in estimation.

Our proposal is to estimate proxy-FA-SVAR models separately for each unit in the panel. For each unit, we specify a SVAR augmented with factors that summarize co-movements  in non-policy variables. The dynamic causal effects of interest are overidentified by internally constructed instruments for the non-policy shocks. These instruments are obtained from auxiliary SVAR models which can be identified through `standard' methods. For example, we use a Blanchard-Perotti-type recursive structure to generate global instruments for regional output shocks. Alternative identification schemes can be applied, depending on the specific context.    
   
We illustrate the practical relevance of the proposed approach by estimating the size and uncertainty of government spending multipliers across Italian regions. The application provides a natural test case for our methodology. On the one hand, the scarcity of official regional fiscal data makes the construction of valid external instruments for policy shocks particularly challenging. On the other hand, the need for region-specific fiscal analysis is especially compelling in the Italian context, where persistent territorial disparities and the long-standing North–South divide make aggregate fiscal multipliers potentially uninformative about local effects. The relatively short time series dimension of the panel calls for caution in the interpretation of results. 

Future research could extend the proposed methodology along two main directions. First, from an econometric perspective, it would be useful to develop bootstrap-based inference procedures that incorporate the uncertainty associated with the different stages of the estimation process. As is typical in FA-SVAR applications, the estimated factors are treated as observed in the subsequent structural analysis under the maintained assumption that the cross-sectional dimension is sufficiently large relative to the time dimension, which is not the scenario that best describes the framework presented in this paper. Thus, a natural refinement would be to account explicitly for the fact that factors are generated regressors, extracted from the same sample on which the proxy-SVAR models are estimated, and to propagate this uncertainty through the construction of the instruments and the estimation of the impulse responses.

Second, from an empirical perspective, the framework could be extended to study regional spillovers. While the present application focuses on the effect of a fiscal shock in a given region on that region's own output, an important issue is whether fiscal shocks also propagate across regional borders. In the Italian case, this would allow one to assess whether a fiscal expansion in region $i$ has measurable effects on economic activity in region $j$. Given the limited time dimension of the data, modelling these spillovers through local projections may be demanding. A feasible alternative would be to estimate `enlarged' proxy-FA-SVAR models for pairs of contiguous regions, or for small regional clusters, thus capturing cross-regional transmission while preserving parsimony.

\clearpage
\bibliographystyle{apalike}
\bibliography{biblio.bib}

	\clearpage

\section*{Appendix A: Factor construction} 
\renewcommand{\theequation}{A.\arabic{equation}}
\renewcommand{\thesection}{A}
\setcounter{equation}{0}
\label{sec:appendix_factor_estimation}
	In this Appendix, we discuss how the common and idiosyncratic factors used in the paper are obtained.  
	
	We assume that, besides a constant $\gamma_i^{(j)}$ and a deterministic time trend $\delta_i^{(j)}t$, each variable $Y_{i,t}^{(j)}, \text{ for } j = 1,\ldots,n \text{ and } i= 1,\ldots,R,$ can be decomposed into a common component, driven by a pervasive factor common to all regions,  and an idiosyncratic component or factor, which can be interpreted as a region-specific driving force. The idiosyncratic factors are allowed to be cross-sectionally mildly correlated. Letting $\tilde{Y}_{i,t}^{(j)} = Y_{i,t}^{(j)}- \gamma_i^{(j)} - \delta_i^{(j)} t$, we have:
	\begin{equation}
		\tilde{Y}_{i,t}^{(j)}  = \lambda_{i}^{(j)}  F_{t}^{(j)} + \eta_{i,t}^{(j)} , \quad i = 1,\ldots, R; \quad j = 1, \ldots, n; \quad t = 1, \ldots, T.
	\end{equation}
where $F_{t}^{(j)} \in \mathcal{D}_t^{j}$ is the common factor, $\lambda_{i}^{(j)}$ is the associated scalar factor loading, and $\eta_{i,t}^{(j)}$ is the idiosyncratic component, orthogonal to $F_{t}^{(j)}$.
	
	Compactly, we can write:
	\begin{equation}
		\tilde{Y}_{t}^{(j)}  = \lambda^{(j)}  F_{t}^{(j)} + \eta_{t}^{(j)} ,  \quad j = 1, \ldots, n; \quad t = 1, \ldots, T.
	\end{equation}
	where $\lambda^{(j)} = (\lambda_1^{(j)}, \ldots, \lambda_R^{(j)})^{\prime}$ is an \emph{R}-dimensional vector of factor loadings and $\eta_{t}^{(j)} = (\eta_{1,t}^{(j)}, \ldots, \eta_{R,t}^{(j)})^{\prime}$ is the \emph{R}-dimensional vector containing the idiosyncratic factors.

	$F_{t}^{o}$ denotes an $ (n-k) \times 1 $ vector of factors from the set of non-policy variables of the panel, $F_t^{o} \in \mathcal{D}_{t}^{Y_o}$;  similarly, $F_{t}^{p}$ denotes a $ k \times 1 $ vector of factors from the set of policy variables of the panel, $F_t^{p} \in \mathcal{D}_{t}^{Y_p}$. In this case, we may consider the models: 
	\begin{align}
		\tilde{Y}_{i,t}^o &= \lambda_{i}^{o} F_{t}^{o} + \eta_{i,t}^{o} , \quad i = 1,\ldots, R;  \quad t = 1, \ldots, T.\\
		\tilde{Y}_{i,t}^p &= \lambda_{i}^{p} F_{t}^{p} + \eta_{i,t}^{p} , \quad i = 1,\ldots, R;  \quad t = 1, \ldots, T.
	\end{align}
	where $\lambda_{i}^{o}$ and $\lambda_{i}^{p}$ are diagonal matrices with elements $\{\lambda_{i}^{(j)} \}_{j\in Y_o}$ and $\{ \lambda_{i}^{(j)}\}_{j\in Y_p}$ on the main diagonals, respectively.
	
	Estimation is performed following \cite{barigozzi2021large}. We estimate $\lambda^{(j)}$ as $\sqrt{R}$ times the eigenvector associated with the largest eigenvalue of the covariance matrix $\bar \Omega^{(j)} := T^{-1}\sum_{t=1}^{T} \bar Y_{t}^{(j)} \bar Y_{t}^{(j)\prime}$, where $\bar Y_{t}^{(j)}$ is the standardized first difference of $Y_{t}^{(j)}$. The estimated factor $\hat F_{t}^{(j)}$ is then obtained by projecting $\hat \lambda^{(j)}$ onto the detrended dataset $ \tilde{Y}_{t}^{(j)}$. Finally, the unit-specific idiosyncratic factors are obtained as $\hat \eta_{t}^{(j)} = \tilde{Y}_{t}^{(j)} - \hat \lambda^{(j)}  \hat F_{t}^{(j)}$.

\end{document}